# Defeasible Modalities

Katarina Britz
Centre for Artificial Intelligence Research
CSIR Meraka Institute and UKZN, South Africa
arina.britz@meraka.org.za

Ivan Varzinczak
Centre for Artificial Intelligence Research
CSIR Meraka Institute and UKZN, South Africa
ivan.varzinczak@meraka.org.za

## ABSTRACT

Nonmonotonic logics are usually characterized by the presence of some notion of 'conditional' that fails monotonicity. Research on nonmonotonic logics is therefore largely concerned with the defeasibility of argument forms and the associated normality (or abnormality) of its constituents. In contrast, defeasible *modes of inference* aim to formalize the defeasible aspects of modal notions such as actions, obligations and knowledge. In this work we enrich the standard possible worlds semantics with a preference ordering on worlds in Kripke models. The resulting family of modal logics allow for the elegant expression of defeasible modalities. We also propose a tableau calculus which is sound and complete with respect to our preferential semantics.

## 1. INTRODUCTION AND MOTIVATION

Defeasible reasoning, as traditionally studied in the literature on nonmonotonic reasoning, has focused mostly on one aspect of defeasibility, namely that of *argument forms*. Such is the case in the approach by Kraus et al. [33, 35], known as the KLM approach, and related frameworks [5, 6, 7, 8, 10, 15, 20, 21]. For instance, in the KLM approach (propositional) defeasible consequence relations $\mid\!\sim$ with a preferential semantics are studied. In this setting, the meaning of a defeasible statement (or a 'conditional', as it is sometimes referred to) of the form $\alpha \mid\!\sim \beta$ is that "all normal $\alpha$-worlds are $\beta$-worlds", leaving it open for $\alpha$-worlds that are, in a sense, exceptional not to satisfy $\beta$. With the theory that has been developed around this notion it becomes possible to cope with exceptionality when performing reasoning.

There are of course many other appealing and equally useful aspects of defeasibility besides that of arguments. These include notions such as typicality [4, 21], concerned with the most typical cases or situations (or even the most typical representatives of a class), and belief plausibility [2], which relates to the most plausible epistemic possibilities held by an agent, amongst others. It turns out that with KLM-style defeasible statements one cannot capture these aspects of defeasibility. This has to do partly with the syntactic restrictions imposed on $\mid\!\sim$, namely no nesting of con-ditionals, but, more fundamentally, it relates to where and how the notion of normality is used in such statements. Indeed, in a KLM defeasible statement $\alpha \mid\!\sim \beta$, the normality spotlight is somewhat put on $\alpha$, as though normality was a property of the premise and not of the conclusion. Whether the situations in which $\beta$ holds are normal or not plays no role in the reasoning that is carried out. In the original KLM framework, normality is also linked to the premise as a whole, rather than its constituents. Technically this meant one could not refer directly to normality of a sentence in the scope of logical operators. This limitation is overcome by taking a (modal) conditional approach *à la* Boutilier [5] — the resulting conditional logics are sufficiently general to allow for the expression of a number of different forms of defeasible reasoning. However, the emphasis remains on the defeasibility of arguments, or of conditionals.

In this paper we investigate a related, but incomparable, notion which we refer to as defeasible *modes of inference* [11].[1] These amount to defeasible versions of the traditional notions of actions, obligations, knowledge and beliefs, to name a few, as studied in modal logics. For instance, in an action context, one can say that normally the outcome of a given action $a$ is $\alpha$. However we may also want to state that the outcome of $a$ is usually (or normally) $\alpha$, which is different from the former statement. To see why, the first statement says that in the most normal worlds, the result of performing the action $a$ is *always* $\alpha$, whereas in the second one it is in the most normal situations resulting from $a$'s execution that $\alpha$ holds — regardless of whether the situation in which the claim is uttered is normal or not.

For a concrete example, assume one arrives at a dark room and wants to toggle the light switch. Exceptionally, the light will not turn on. This can be either because the light bulb is blown (the current situation is abnormal) or because an overcharge was caused while switching the light (the action behaves abnormally). In the former case, the normality of the situation, or state, before the action is assessed, whereas in the latter the relative normality of the situation is assessed against all possible outcomes. Here we are interested in the formalization of the latter type of statement, where it becomes important to shift the notion of normality from the premise of an inference to the effect of an action, and, importantly, use it in the scope of other logical constructors.

Our next example concerns obligations and weaker versions thereof. There is a subtle difference between stating that, from the perspective of any normal situation, rhino poaching ought to carry a minimum prison sentence of 10 years, and stating that, from any perspective, the minimum sentence for rhino poaching normally ought to be 10 years. The shift in focus is again from normality of the present world, to relative normality amongst possible worlds. In the former statement, an abnormal present world would render the obligation unenforceable, whereas in the latter statement, the obligation is applicable in all relatively normal accessible worlds. We contend that the informal notion of 'normal, reasonable obligations' is more accurately modeled as defeasible modalities than as conditional statements.

---

[1] The present paper extends and refines the preliminary proposal which was presented at the 14th International Workshop on Non-Monotonic Reasoning (NMR).

Scenarios such as the ones depicted above require an ability to talk about the normality of effects of an action, relative normality of obligations, and so on. While existing modal treatments of preferential reasoning can express preferential semantics syntactically as modalities [5, 6, 20], they do not suffice to express defeasible modes of inference as described above. The ability to capture precisely these forms of defeasibility remains a fundamental challenge in the definition of a coherent theory of defeasible reasoning. At present we can formalize only the first type of statements above by, e.g. stating $\top \mid\!\sim \Box\alpha$ in Britz et al.'s extension of preferential reasoning to modal languages [8, 10]. (As we shall see later in the paper, both Boutilier's [5] and Booth et al.'s [4] approaches also have to be enriched in order to capture the forms of defeasibility we are interested in here.)

In this paper we make the first steps towards filling this gap by introducing (non-standard) modal operators allowing us to talk about relative normality in accessible worlds. With our defeasible versions of modalities, we can make statements of the form "$\alpha$ holds in all of the relatively normal accessible worlds", thereby capturing defeasibility of what is 'expected' in target worlds. This notion of defeasibility in a modality meets a variety of applications in Artificial Intelligence, ranging from reasoning about actions to deontic and epistemic reasoning. For instance, a defeasible-action operator allows us to make statements of the form $\boxdot_a \alpha$, which we read as "$\alpha$ is a normal necessary effect of $a$" (i.e., necessary in the most normal of $a$'s outcomes), and with defeasible-obligation operators one can state formulae such as $\boxdot_A \alpha$, read as "$\alpha$ is a normal obligation of agent $A$".

These operators are defined within the context of a general preferential modal semantics obtained by enriching the standard possible worlds semantics with a preference order. The main difference between the approach we propose here and that of Boutilier [5] is in whether the underlying preference ordering alters the meaning of modalities or not. Boutilier's conditional is defined directly from a preference order in a bi-modal language, but the meanings of any additional, independently axiomatized, modalities are not influenced by the preference order. Our defeasible modalities correspond to a modification of the other modalities using the preference relation. Also, in contrast with the plausibility models of Baltag and Smets [2], the preference order we consider here does not define an agent's knowledge or beliefs. Rather, it is part of the semantics of the background ontology described by the theory or knowledge base at hand. As such, it informs the meaning of defeasible actions, which can fail in their outcome, or defeasible obligations, which may not hold in exceptional accessible worlds, in that it alters the classical semantics of these modalities. This allows for the definition of a family of modal logics in which defeasible modes of inference can be expressed, and which can be integrated with existing $\mid\!\sim$-based nonmonotonic modal logics [8, 10].

The remainder of the present paper is structured as follows: After setting up the notation and terminology that we shall follow in this paper (Section 2), we revisit Britz et al.'s preferential semantics for modal logic [8, 10] (Section 3) by proposing a simplified version thereof. In Section 4 we present a logic enriched with defeasible modalities allowing for the formalization of defeasible versions of modes of inference. In Section 5 we present a detailed example illustrating the application of our constructions in an action context. Following that, we define a tableau system for the corresponding logic that we show to be sound and complete with respect to our preferential semantics (Section 6). In Section 7 we assess $\mid\!\sim$-statements in our richer language. After a discussion of and comparison with related work (Section 8), we conclude with some comments and directions for further investigation. All the proofs of our results can be found in the Appendix.

## 2. MODAL LOGIC

We assume the reader is familiar with modal logic [14]. The purpose of this section is to make explicit the terminology and notation we shall use.

Here we work within a set of *atomic propositions* $\mathcal{P}$, using the logical connectives $\wedge$ (conjunction), $\neg$ (negation), and a set of modal operators $\Box_i$, $1 \leq i \leq n$. (In later sections we shall adopt a richer language.) We assume that the underlying multimodal logic is independently axiomatized (i.e., the logic is a fusion and there is no interaction between the modal operators [32]). Propositions are denoted by $p, q, \ldots$, and formulae by $\alpha, \beta, \ldots$, constructed in the usual way according to the rule: $\alpha ::= p \mid \neg \alpha \mid \alpha \wedge \alpha \mid \Box_i \alpha$. All the other truth functional connectives ($\vee, \rightarrow, \leftrightarrow, \ldots$) are defined in terms of $\neg$ and $\wedge$ in the usual way. Given $\Box_i$, $1 \leq i \leq n$, with $\Diamond_i$ we denote its *dual* modal operator, i.e., for any $\alpha$, $\Diamond_i \alpha \equiv_{\text{def}} \neg \Box_i \neg \alpha$. We use $\top$ as an abbreviation for $p \vee \neg p$, and $\bot$ as an abbreviation for $p \wedge \neg p$, for some $p \in \mathcal{P}$.

With $\mathcal{L}$ we denote the set of all formulae of the modal language. The semantics is the standard possible-worlds one:

DEFINITION 1. *A Kripke model is a tuple $\mathscr{M} = \langle W, R, V \rangle$ where $W$ is a (non-empty) set of possible worlds, $R = \langle R_1, \ldots, R_n \rangle$, where each $R_i \subseteq W \times W$ is an accessibility relation on $W$, $1 \leq i \leq n$, and $V: W \times \mathcal{P} \longrightarrow \{0, 1\}$ is a valuation function.*

Satisfaction of formulae with respect to possible worlds in a Kripke model is defined in the usual way:

DEFINITION 2. *Let $\mathscr{M} = \langle W, R, V \rangle$ and $w \in W$:*

- $\mathscr{M}, w \Vdash p$ *if and only if* $V(w, p) = 1$;
- $\mathscr{M}, w \Vdash \neg \alpha$ *if and only if* $\mathscr{M}, w \not\Vdash \alpha$;
- $\mathscr{M}, w \Vdash \alpha \wedge \beta$ *if and only if* $\mathscr{M}, w \Vdash \alpha$ *and* $\mathscr{M}, w \Vdash \beta$;
- $\mathscr{M}, w \Vdash \Box_i \alpha$ *if and only if* $\mathscr{M}, w' \Vdash \alpha$ *for all $w'$ such that* $(w, w') \in R_i$.

Given $\alpha \in \mathcal{L}$ and $\mathscr{M} = \langle W, R, V \rangle$, we say that $\mathscr{M}$ *satisfies* $\alpha$ if there is at least one world $w \in W$ such that $\mathscr{M}, w \Vdash \alpha$. We say that $\mathscr{M}$ is a *model of* $\alpha$ (alias $\alpha$ is *true* in $\mathscr{M}$), denoted $\mathscr{M} \Vdash \alpha$, if $\mathscr{M}, w \Vdash \alpha$ for every world $w \in W$. Given a class of models $\mathcal{M}$, we say that $\alpha$ is *valid* in $\mathcal{M}$ if every Kripke model $\mathscr{M} \in \mathcal{M}$ is a model of $\alpha$.

Here we shall assume the system of normal modal logic K, of which all the other normal modal logics are extensions. Semantically, K is characterized by the class of all Kripke models (Definition 1). We say that $\alpha$ *locally entails* $\beta$ in the system K (denoted $\alpha \models \beta$) if for every K-model $\mathscr{M}$ and every $w$ in $\mathscr{M}$, $\mathscr{M}, w \Vdash \alpha$ implies $\mathscr{M}, w \Vdash \beta$.

Syntactically, K corresponds to the smallest set of sentences containing all propositional tautologies, all instances of the axiom schema $K : \Box_i(\alpha \to \beta) \to (\Box_i \alpha \to \Box_i \beta)$, $1 \leq i \leq n$, and closed under the *rule of necessitation RN*: $\alpha/\Box_i \alpha$, $1 \leq i \leq n$.

## 3. MODAL PREFERENTIAL SEMANTICS

In this section we modify the constructions for preferential reasoning in modal logic as studied by Britz et al. [8, 10]. We do so by enriching standard Kripke models with preference relations, instead of placing an ordering on states which are labeled with *pointed* Kripke models. Our starting point is therefore similar to the CT4O models of Boutilier [5] and the plausibility models of Baltag and Smets [2].

DEFINITION 3. *A preferential Kripke model is a tuple $\mathscr{P} := \langle W, R, V, \prec \rangle$ where $W$ is a (non-empty) set of possible worlds, $R = \langle R_1, \ldots, R_n \rangle$, where each $R_i \subseteq W \times W$ is an accessibility relation on $W$, $1 \leq i \leq n$, $V : W \times \mathcal{P} \longrightarrow \{0, 1\}$ is a valuation function, and $\prec \subseteq W \times W$ is a co-Noetherian strict partial order on $W$, i.e., $\prec$ is irreflexive, transitive and well-founded.*[2]

Given a preferential Kripke model $\mathscr{P} = \langle W, R, V, \prec \rangle$, we refer to $\mathscr{M} := \langle W, R, V \rangle$ as its associated standard Kripke model. If $\mathscr{P} = \langle W, R, V, \prec \rangle$ is a preferential Kripke model and $\alpha \in \mathcal{L}$, then with $[\![\alpha]\!] := \{w \in W \mid \mathscr{M}, w \Vdash \alpha$, where $\mathscr{M} = \langle W, R, V \rangle \}$ we denote the set of possible worlds satisfying $\alpha$ ($\alpha$-worlds for short).

DEFINITION 4. *Let $\mathscr{P} = \langle W, R, V, \prec \rangle$ and let $W' \subseteq W$. With $\min_\prec W'$ we denote the minimal elements of $W'$ with respect to $\prec$, i.e., $\min_\prec W' := \{w \in W' \mid$ there is no $w' \in W'$ such that $w' \prec w\}$.*

The intuition behind the preference relation $\prec$ in a preferential Kripke model $\mathscr{P}$ is that worlds lower down in the order are *more preferred* (or *more normal* [4, 5]) than those higher up. Note that the preference relation in a preferential Kripke model, although a binary relation on $W$, is not to be seen as an accessibility relation. Indeed, the $\prec$-component in a preferential Kripke model has no counterpart in the syntax as each accessibility relation has.

As an example, Figure 1 below depicts the preferential Kripke model $\mathscr{P} = \langle W, R, V, \prec \rangle$, where $W = \{w_i \mid 1 \leq i \leq 5\}$, $R = \langle R_\Box \rangle$, with $R_\Box = \{(w_1, w_2), (w_1, w_4), (w_2, w_3), (w_2, w_5), (w_3, w_2), (w_4, w_5), (w_5, w_4)\}$, represented by the solid arrows in the picture, $V$ is the obvious valuation function (in our pictorial representations of models we interpret absence of an atom as the atom being false in the respective world), and $\prec$ is the transitive closure of $\{(w_1, w_2), (w_1, w_3), (w_2, w_4), (w_3, w_4), (w_4, w_5)\}$, represented by the dashed arrows in the picture. (Note the direction of the dashed arrows, which point from less preferred to more preferred worlds.)

---

[2]This implies the *smoothness condition* in Kraus et al.'s terminology [33], which basically says that $\prec$ has no infinitely descending chains. Even though well-foundedness is stronger than smoothness, here we prefer to stick to the term that is more broadly known outside the nonmonotonic reasoning circle.

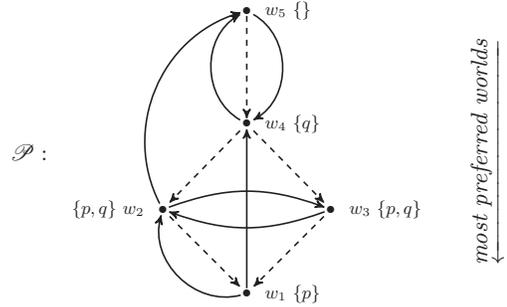

**Figure 1: A preferential Kripke model for $\mathcal{P} = \{p, q\}$ and a single modality.**

Given $\mathscr{P} = \langle W, R, V, \prec \rangle$ and $\alpha \in \mathcal{L}$, $\alpha$ is *satisfiable* in $\mathscr{P}$ if $[\![\alpha]\!] \neq \emptyset$, otherwise $\alpha$ is *unsatisfiable* in $\mathscr{P}$. We say that $\alpha$ is *true* in $\mathscr{P}$ (denoted $\mathscr{P} \Vdash \alpha$) if $[\![\alpha]\!] = W$. It is easy to see that the addition of the $\prec$-component preserves the truth of all (classical) modal formulae that are true in the remaining Kripke structure:

LEMMA 1. *Let $\alpha \in \mathcal{L}$ (i.e., $\alpha$ is a classical modal formula). Let $\mathscr{P} = \langle W, R, V, \prec \rangle$ be a preferential Kripke model and $\mathscr{M} = \langle W, R, V \rangle$ its associated standard Kripke model. Then $\mathscr{P} \Vdash \alpha$ if and only if $\mathscr{M} \Vdash \alpha$.*

PROOF. See Appendix A.1. $\square$

We can define *classes* of preferential Kripke models in the same way we do in the classical case. For instance, we can talk about the class of reflexive preferential Kripke models, in which the $R$-components are reflexive. We say that $\alpha$ is *valid* in the class $\mathcal{M}$ of preferential Kripke models if and only if $\alpha$ is true in every $\mathscr{P} \in \mathcal{M}$. Therefore, the following result is an immediate consequence of Lemma 1:

COROLLARY 1. *A modal formula $\alpha$ is valid in the class $\mathcal{M}$ of preferential Kripke models if and only if it is valid in the corresponding class of Kripke models.*

## 4. PREFERENCE-BASED MODALITIES

Recalling our discussion in the Introduction, we want to be able to state that a given sentence holds in *all* the relatively normal worlds that are accessible. This leads us to the definition of a 'weaker' version of the $\Box$ modalities. Through them we are then able to single out those normal situations that one cannot grasp via the classical $\Box$ modalities. Similarly, we want to be able to state that a given sentence holds in *at least one* relatively normal accessible world. This leads us to the definition of a stronger version of $\Diamond$, which may be read as *distinct* possibility.

We define a more expressive language than $\mathcal{L}$ by extending our modal language with a family of defeasible modal operators $\boxdot_i$ and $\diamondsuit_i$, $1 \leq i \leq n$ (called, respectively, the 'flag' and the 'flame'), where $n$ is the number of classical modalities in the language. The formulae of the extended language are then recursively defined by:

$$\alpha ::= p \mid \neg\alpha \mid \alpha \wedge \alpha \mid \Box_i\alpha \mid \mathbin{\!\!\vcenter{\hbox{$\sim$}}\mkern-8mu{\vcenter{\hbox{$\sim$}}}\!\!}_i\alpha \mid \Diamond_i\alpha$$

(As before, the other connectives are defined in terms of $\neg$ and $\wedge$ in the usual way, and $\top$ and $\bot$ are seen as abbreviations. It turns out that each $\Diamond_i$ too is the dual of $\mathbin{\!\!\vcenter{\hbox{$\sim$}}\mkern-8mu{\vcenter{\hbox{$\sim$}}}\!\!}_i$, as we shall see below.) With $\widetilde{\mathcal{L}}$ we denote the set of all formulae of such a richer language.

The semantics of $\widetilde{\mathcal{L}}$ is in terms of our preferential Kripke models (see Definition 3). As before, given $\alpha \in \widetilde{\mathcal{L}}$ and a preferential Kripke model $\mathscr{P} = \langle W, R, V, \prec \rangle$, with $[\![\alpha]\!]$ we denote the set of elements of $W$ satisfying $\alpha$.

DEFINITION 5. *Let $\mathscr{P} = \langle W, R, V, \prec \rangle$ be a preferential Kripke model. Then:*

- $[\![\mathbin{\!\!\vcenter{\hbox{$\sim$}}\mkern-8mu{\vcenter{\hbox{$\sim$}}}\!\!}_i\alpha]\!] := \{w \in W \mid \min_\prec R_i(w) \subseteq [\![\alpha]\!]\}$;
- $[\![\Diamond_i\alpha]\!] := \{w \in W \mid \min_\prec R_i(w) \cap [\![\alpha]\!] \neq \emptyset\}$.

The intuition behind a sentence like $\mathbin{\!\!\vcenter{\hbox{$\sim$}}\mkern-8mu{\vcenter{\hbox{$\sim$}}}\!\!}_i\alpha$ is that $\alpha$ holds in the most 'normal' of $R_i$-accessible worlds. $\Diamond_i\alpha$ intuitively says that $\alpha$ holds in at least one such relatively normal accessible world. To give a simple example (a more elaborated one is given in Section 5), if toggle denotes the action of toggling the light switch and light the proposition "the light is on", with the formula $\neg$light $\to \mathbin{\!\!\vcenter{\hbox{$\sim$}}\mkern-8mu{\vcenter{\hbox{$\sim$}}}\!\!}_{\mathsf{toggle}}$light we formalize the example from the Introduction.

As mentioned before, in our enriched language the preference relation is not explicit in the syntax. The meaning of the new modalities is informed by the preference relation, which nevertheless remains tacit outside the realm of defeasible modalities. This stands in contrast to the approaches of Baltag and Smets [2], Boutilier [5], Britz et al. [6] and Giordano et al. [20], which cast the preference relation as an extra modality in the language. From a knowledge representation perspective, our approach has the advantage of hiding some complex aspects of the semantics from the user (e.g. a knowledge engineer who will write down sentences in an agent's knowledge base).

The notions of satisfaction in a preferential Kripke model, truth (in a model) and validity (in a class of preferential Kripke models) are extended to formulae with defeasible modalities in the obvious way.

We observe that, like in the classical (i.e., non-defeasible) case, the defeasible modal operators $\mathbin{\!\!\vcenter{\hbox{$\sim$}}\mkern-8mu{\vcenter{\hbox{$\sim$}}}\!\!}$ and $\Diamond$ are the dual of each other:

$$\models \mathbin{\!\!\vcenter{\hbox{$\sim$}}\mkern-8mu{\vcenter{\hbox{$\sim$}}}\!\!}_i\alpha \leftrightarrow \neg\Diamond_i\neg\alpha \tag{1}$$

The following validities are also easy to verify:

$$\models \mathbin{\!\!\vcenter{\hbox{$\sim$}}\mkern-8mu{\vcenter{\hbox{$\sim$}}}\!\!}_i\bot \leftrightarrow \Box_i\bot \qquad \models \Diamond_i\top \leftrightarrow \Diamond_i\top$$
$$\models \mathbin{\!\!\vcenter{\hbox{$\sim$}}\mkern-8mu{\vcenter{\hbox{$\sim$}}}\!\!}_i\top \leftrightarrow \top \qquad \models \Diamond_i\bot \leftrightarrow \bot$$

The following is the $\mathbin{\!\!\vcenter{\hbox{$\sim$}}\mkern-8mu{\vcenter{\hbox{$\sim$}}}\!\!}$-version of Axiom Schema $K$.

$$(\widetilde{K}) \quad \models \mathbin{\!\!\vcenter{\hbox{$\sim$}}\mkern-8mu{\vcenter{\hbox{$\sim$}}}\!\!}_i(\alpha \to \beta) \to (\mathbin{\!\!\vcenter{\hbox{$\sim$}}\mkern-8mu{\vcenter{\hbox{$\sim$}}}\!\!}_i\alpha \to \mathbin{\!\!\vcenter{\hbox{$\sim$}}\mkern-8mu{\vcenter{\hbox{$\sim$}}}\!\!}_i\beta) \tag{2}$$

The validity below is easy to verify:

$$(\widetilde{R}) \quad \models \mathbin{\!\!\vcenter{\hbox{$\sim$}}\mkern-8mu{\vcenter{\hbox{$\sim$}}}\!\!}_i(\alpha \wedge \beta) \leftrightarrow (\mathbin{\!\!\vcenter{\hbox{$\sim$}}\mkern-8mu{\vcenter{\hbox{$\sim$}}}\!\!}_i\alpha \wedge \mathbin{\!\!\vcenter{\hbox{$\sim$}}\mkern-8mu{\vcenter{\hbox{$\sim$}}}\!\!}_i\beta) \tag{3}$$

We also have $\models (\mathbin{\!\!\vcenter{\hbox{$\sim$}}\mkern-8mu{\vcenter{\hbox{$\sim$}}}\!\!}_i\alpha \vee \mathbin{\!\!\vcenter{\hbox{$\sim$}}\mkern-8mu{\vcenter{\hbox{$\sim$}}}\!\!}_i\beta) \to \mathbin{\!\!\vcenter{\hbox{$\sim$}}\mkern-8mu{\vcenter{\hbox{$\sim$}}}\!\!}_i(\alpha \vee \beta)$, but not the converse, as can easily be checked.

The following validity is an immediate consequence of our preferential semantics:

$$(N) \models \Box_i\alpha \to \mathbin{\!\!\vcenter{\hbox{$\sim$}}\mkern-8mu{\vcenter{\hbox{$\sim$}}}\!\!}_i\alpha \tag{4}$$

Intuitively, given $i = 1, \ldots, n$, where $n$ is the number of modalities in the language, we want $\Box_i$ and $\mathbin{\!\!\vcenter{\hbox{$\sim$}}\mkern-8mu{\vcenter{\hbox{$\sim$}}}\!\!}_i$ to be 'tied' together in so far as one is the defeasible (or the 'hard') version of the other. Schema $N$ is in line with the commonly accepted principle that whatever is classically the case is also defeasibly so.[3]

From duality of $\Diamond$ and $\mathbin{\!\!\vcenter{\hbox{$\sim$}}\mkern-8mu{\vcenter{\hbox{$\sim$}}}\!\!}$ and contraposition of $N$ we get:

$$\models \Diamond_i\alpha \to \Diamond_i\alpha \tag{5}$$

It can easily be checked that in our preferential semantics, the standard rule of necessitation $RN : \alpha/\Box_i\alpha$ holds. The following rule of *normal necessitation* ($RNN$) follows from $RN$ together with Schema $N$ in (4) above:

$$(RNN) \; \frac{\alpha}{\mathbin{\!\!\vcenter{\hbox{$\sim$}}\mkern-8mu{\vcenter{\hbox{$\sim$}}}\!\!}_i\alpha} \tag{6}$$

From satisfaction of (1), (2) and (3), one can see that the logic of our defeasible modalities shares properties commonly characterizing the so-called *normal* modal logics [14]. In particular, we have that the following rule holds:

$$(NRK) \; \frac{(\alpha_1 \wedge \ldots \wedge \alpha_n) \to \beta}{(\mathbin{\!\!\vcenter{\hbox{$\sim$}}\mkern-8mu{\vcenter{\hbox{$\sim$}}}\!\!}_i\alpha_1 \wedge \ldots \wedge \mathbin{\!\!\vcenter{\hbox{$\sim$}}\mkern-8mu{\vcenter{\hbox{$\sim$}}}\!\!}_i\alpha_n) \to \mathbin{\!\!\vcenter{\hbox{$\sim$}}\mkern-8mu{\vcenter{\hbox{$\sim$}}}\!\!}_i\beta} \; (n \geq 0) \tag{7}$$

The observant reader would have noticed that we assume we have as many defeasible modalities as we have classical ones. That is, for each $\Box_i$, a corresponding $\mathbin{\!\!\vcenter{\hbox{$\sim$}}\mkern-8mu{\vcenter{\hbox{$\sim$}}}\!\!}_i$ (its defeasible version) is assumed. Moreover they are both linked together via Schema $N$ in (4). In principle, from a technical point of view, nothing precludes us from having defeasible modalities with no corresponding classical version or the other way round. The latter is easily dealt with by simply not having $\mathbin{\!\!\vcenter{\hbox{$\sim$}}\mkern-8mu{\vcenter{\hbox{$\sim$}}}\!\!}_i$ for some $i$ for which $\Box_i$ is present in the language. The former case, on the other hand, would require an elaboration of the semantics as satisfiability of $\mathbin{\!\!\vcenter{\hbox{$\sim$}}\mkern-8mu{\vcenter{\hbox{$\sim$}}}\!\!}$-formulae calls upon the accessibility relation $R_i$, associated with the $\Box_i$-modality.

The dependency between each (classical) modality and its defeasible counterpart is defined by a (fixed) preference order on worlds in the model. We do not have a Hilbert-style axiomatization of this dependency yet. What is certain is that such an axiomatization would require casting the preference order as a modality, in order to axiomatize the relationship between $\mathbin{\!\!\vcenter{\hbox{$\sim$}}\mkern-8mu{\vcenter{\hbox{$\sim$}}}\!\!}_i$, $\Diamond_i$ and the preference order $\prec$, for each $i$. To this end, we may use, for example, the modal axiomatization of the preference order of Britz et al. [6], or one of Boutilier's modal systems [5]. Such an axiomatization is possible at the expense of moving to a more expressive language (see the remark below Definition 5 above and also the discussion in Section 8). Nevertheless, from a computational logic point of view, we shall suffice with the definition of a tableau-based decision procedure, which will be presented in Section 6.

We also observe that in order for us to capture the semantics of $\widetilde{\mathcal{L}}$ in standard conditional logics [14] we would require the addition of a preference relation on worlds, all standard modalities we want to work with and a suitably defined conditional for each modality in the language. Our contention here is that this route would hardly simplify matters.

---

[3]Similarly to what happens in KLM consequence relations ($\alpha \models \beta$ implies $\alpha \mathrel{\mid\!\sim} \beta$) [33] and in defeasible subsumption relations ($C \sqsubseteq D$ implies $C \mathrel{\underset{\sim}{\sqsubset}} D$) [9].

From the perspective of knowledge representation and reasoning, it becomes important to address the question of what it means for an $\widetilde{\mathcal{L}}$-sentence to be *entailed* from an $\widetilde{\mathcal{L}}$-knowledge base.

An $\widetilde{\mathcal{L}}$-*knowledge base* is a (possibly infinite) set $\mathcal{K} \subseteq \widetilde{\mathcal{L}}$. Given a preferential Kripke model $\mathscr{P}$, we extend the notion of satisfaction to knowledge bases in the obvious way: $\mathscr{P} \Vdash \mathcal{K}$ if and only if $\mathscr{P} \Vdash \alpha$ for every $\alpha \in \mathcal{K}$.

DEFINITION 6. *Let $\mathcal{K} \subseteq \widetilde{\mathcal{L}}$ and let $\alpha \in \widetilde{\mathcal{L}}$. We say that $\mathcal{K}$ (globally) entails $\alpha$ in the class $\mathcal{M}$ of preferential Kripke models (denoted $\mathcal{K} \models \alpha$) if and only if for every $\mathscr{P} \in \mathcal{M}$, if $\mathscr{P} \Vdash \mathcal{K}$, then $\mathscr{P} \Vdash \alpha$.*

Given this notion of entailment, its associated *consequence relation* is defined as follows:

$$Cn(\mathcal{K}) \equiv_{\text{def}} \{\alpha \mid \mathcal{K} \models \alpha\} \qquad (8)$$

It can be checked that the consequence relation $Cn(\cdot)$ as defined in (8) above is a Tarskian consequence relation:

THEOREM 1. *Let $Cn(\cdot)$ be a consequence relation defined in terms of preferential entailment. Then $Cn(\cdot)$ satisfies the following properties:*

- $\mathcal{K} \subseteq Cn(\mathcal{K})$ *(Inclusion)*
- $Cn(\mathcal{K}) = Cn(Cn(\mathcal{K}))$ *(Idempotency)*
- *If $\mathcal{K}_1 \subseteq \mathcal{K}_2$, then $Cn(\mathcal{K}_1) \subseteq Cn(\mathcal{K}_2)$* *(Monotonicity)*

PROOF. See Appendix A.2. □

That is, in spite of the defeasibility features of $\mathrel{\reflectbox{$\leadsto$}}$, we end up with a logic that is *monotonic* (at the entailment level).

## 5. AN APPLICATION EXAMPLE

Let us assume the following simple scenario depicting a nuclear power-plant [8]. In a particular power station there is an atomic pile and a cooling system, both of which can be either on or off. A surveillance agent is in charge of detecting hazardous situations so that the human controller can prevent the plant from malfunctioning (Figure 2).

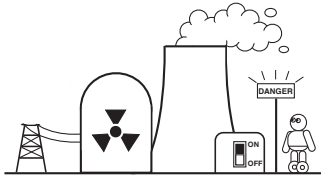

**Figure 2: The power plant and its surveillance agent.**

In what follows we shall illustrate our constructions from previous sections in reasoning about action using the aforementioned scenario.

We find in the AI literature a fair number of modal-based formalisms for reasoning about actions and change [12, 13, 16, 18, 29, 38, 41, 42, 43]. These are essentially variants of the modal logic K we presented in Section 2. Modal operators are determined by a (finite) set of *actions* $\mathcal{A} = \{a_1, \ldots, a_n\}$: For each $a \in \mathcal{A}$, there is associated a modal operator $\Box_a$. Given a Kripke model, $R_a \subseteq W \times W$ is therefore meant to represent possible executions of an (ontic) action $a$ at specific worlds $w \in W$, i.e., $R_a$ is the specification of $a$'s behavior in a transition system. Hence, whenever $(w, w') \in R_a$, $w'$ is a *possible outcome* of doing $a$ in $w$. Formulae of the form $\Box_a \alpha$ are used to specify the *effects* of actions and they are read "after *every* execution of action $a$, the formula $\alpha$ holds". The operator $\Diamond_a$ is mostly used to specify the *executability* of actions: $\Diamond_a \top$ reads "there is a possible execution of action $a$".

In our nuclear power plant example, let $\mathcal{P} = \{p, c, h\}$ be a set of propositions, where p stands for "the atomic pile is on", c for "the cooling system is on", and h for "hazardous situation". Moreover, let $\mathcal{A} = \{f, m\}$ be a set of atomic actions, where f stands for "flipping the pile switch", and m for (occurrence of) "a malfunction".

We first construct a preferential Kripke model (Definition 3) in which to check the satisfiability and truth of a few sentences. (The purpose is to illustrate the semantics of our notion of defeasibility in an action context rather than to present a comprehensive modeling of the nuclear power plant scenario.)

Let $\mathscr{P} = \langle W, R, V, \prec \rangle$ be the preferential Kripke model depicted in Figure 3, where $W = \{w_i \mid 1 \leq i \leq 4\}$, $R = \langle R_f, R_m \rangle$, with $R_f = \{(w_1, w_2), (w_2, w_1), (w_3, w_1), (w_3, w_4), (w_4, w_2), (w_4, w_4)\}$ and $R_m = \{(w_4, w_3), (w_4, w_4)\}$, $V$ is the obvious valuation function, and $\prec$ is the transitive closure of $\{(w_1, w_2), (w_2, w_3), (w_3, w_4)\}$, i.e., of the relation represented by the dashed arrows in the picture. (Note again the direction of $\prec$ from less to more normal worlds.)

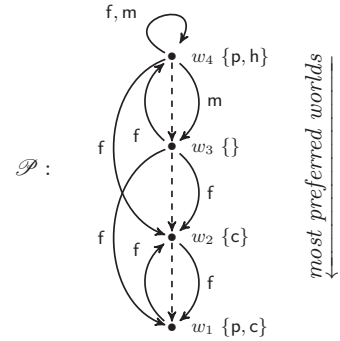

**Figure 3: Preferential Kripke model for the power plant scenario.**

The rationale of this partial order is as follows: The utility company selling the electricity generated by the power plant tries as far as possible to keep both the pile and the cooling system on, ensuring that the pile can easily be switched off (world $w_1$); sometimes the company has to switch the pile off for maintenance but then tries to keep the cooler running, because turning the pile on again would not cause a fault in the cooling system (world $w_2$); more rarely the company needs to switch off both the pile and the cooler, e.g. when the latter needs maintenance (world $w_3$); and, finally, only in very exceptional situations would the pile be on while the cooler is off, e.g. during a serious malfunction (world $w_4$).

In the preferential model $\mathscr{P}$ depicted above, one can check that $\mathscr{P} \Vdash (p \wedge \neg c) \leftrightarrow h$, i.e., $[\![(p \wedge \neg c) \leftrightarrow h]\!] = W$. Also, $w_4 \in [\![h \wedge \Diamond_f \neg h]\!]$: at $w_4$ we have a hazardous situation, but it is possible to switch the pile off having as a normal effect a safe condition. We have that $w_1$ satisfies $\boxtimes_m \bot$: at $w_1$ a malfunction cannot occur (which is not true of $w_4$). In $\mathscr{P}$ we have $\mathscr{P} \Vdash \neg p \to \boxtimes_f p$ (the normal outcome of switching the pile on is it being on), but $\mathscr{P} \nVdash \neg p \to \Box_f p$ (see world $w_3$). We also have $\mathscr{P} \Vdash c \to \boxtimes_f \neg h$ (if the cooler is on, the normal result of switching the pile is a safe situation). Finally we also have $\mathscr{P} \Vdash h \to \Diamond_m \top$: in any hazardous situation a meltdown is a distinct possibility — but fortunately $\mathscr{P} \Vdash \Diamond_f \neg h$: from every world it is possible to return to a non-hazardous world.

So far we have illustrated the preferential semantics of $\widetilde{\mathcal{L}}$-statements using a specific preferential Kripke model. In a knowledge representation context, though, we are interested in preferential entailment from an $\widetilde{\mathcal{L}}$-theory or knowledge base. The latter determines the preferential models that are permissible from the standpoint of the knowledge engineer. To illustrate this, consider the following $\widetilde{\mathcal{L}}$-knowledge base:

$$\mathcal{K} = \left\{ \begin{array}{l} (p \land \neg c) \leftrightarrow h,\ h \to \Diamond_m \top, \\ p \to \boxtimes_f \neg p,\ c \to \boxtimes_f c,\ \Diamond_f \neg h \end{array} \right\}$$

$\mathcal{K}$ basically says that "a hazardous situation is one in which the pile is on and the cooler off", "in a hazardous situation a malfunction is distinctly possible", "if the pile is on, then flipping its switch normally switches it off", "if the cooler is on, then switching the pile normally does not affect it" and "it is always possible to flip the pile switch". (Note that all the formulae in $\mathcal{K}$ are true in the preferential model $\mathscr{P}$ of Figure 3 above.) We can then conclude $\mathcal{K} \models p \to \boxtimes_f \neg h$, $\mathcal{K} \models \boxtimes_m \bot \to (\neg p \lor c)$ and $\mathcal{K} \models (p \lor c) \to \boxtimes_f \neg h$, using the sound $\widetilde{\mathcal{L}}$-inference rules and validities presented in the previous section.

## 6. TABLEAU SYSTEM

In this section we present a simple tableau calculus for defeasible modalities based on labeled formulae and on explicit accessibility relations [22].[4] As we shall see, it also makes use of an auxiliary structure of which the intention is to build a preference relation on possible worlds. (For a discussion on the differences between our tableau method and the one by Giordano et al. [20], see end of Section 8.)

DEFINITION 7. *If $n \in \mathbb{N}$ and $\alpha \in \widetilde{\mathcal{L}}$, then $n :: \alpha$ is a labeled formula.*

In a labeled formula $n :: \alpha$, $n$ is the *label*. (As we shall see, informally, the idea is that the label stands for some possible world in a Kripke model.)

Let $mod(\widetilde{\mathcal{L}})$ denote the set of all *classical modalities* of $\widetilde{\mathcal{L}}$. (Remember our assumption that we have as many defeasible modalities as we have classical ones and that, for a given $i$, both $\Box_i$ and $\boxtimes_i$ depend on the same $R_i$.)

DEFINITION 8. *A skeleton is a function $\Sigma : mod(\widetilde{\mathcal{L}}) \longrightarrow 2^{\mathbb{N} \times \mathbb{N}}$.*

Informally, a skeleton maps modalities in the language to accessibility relations on possible worlds.

DEFINITION 9. *A preference relation $\prec$ is a binary relation on $\mathbb{N}$.*

---
[4]Our exposition here follows that given by Varzinczak [37] and Castilho et al. [12, 13].

As alluded to above, $\prec$ is meant to capture a preference relation on possible worlds. As we shall see below, like $\Sigma$, $\prec$ is built cumulatively through successive applications of the tableau rules we shall introduce.

DEFINITION 10. *A branch is a tuple $\langle \mathcal{S}, \Sigma, \prec \rangle$, where $\mathcal{S}$ is a set of labeled formulae, $\Sigma$ is a skeleton and $\prec$ is a preference relation.*

DEFINITION 11. *A tableau rule is a rule of the form:*

$$\rho \ \frac{\mathcal{N}\ ;\ \Gamma}{\mathcal{D}_1\ ;\ \Gamma'_1\ |\ \ldots\ |\ \mathcal{D}_k\ ;\ \Gamma'_k}$$

*where $\mathcal{N}; \Gamma$ is the* numerator *and $\mathcal{D}_1\ ;\ \Gamma'_1\ |\ \ldots\ |\ \mathcal{D}_k\ ;\ \Gamma'_k$ is the* denominator.

Given a rule $\rho$, $\mathcal{N}$ represents one or more labeled formulae, called the *main formulae* of the rule, separated by ','. $\Gamma$ stands for any additional *condition* (on $\Sigma$ or $\prec$) that must be satisfied for the rule to be applicable. In the denominator, each $\mathcal{D}_i$, $1 \leq i \leq k$, has one or more labeled formulae, whereas each $\Gamma'_i$ is a condition to be satisfied *after* the application of the rule (e.g. changes in the skeleton $\Sigma$ or in the relation $\prec$). The symbol '$|$' indicates the occurrence of a *split* in the branch.

Figure 4 presents the set of tableau rules for $\widetilde{\mathcal{L}}$. In the rules we abbreviate $(n, n') \in \Sigma(i)$ as $n \xrightarrow{i} n'$, and $n' \in \Sigma(i)(n)$ as $n' \in \Sigma_i(n)$. Finally, with $n'^\star, n''^\star, \ldots$ we denote labels that have not been used before. We say that a rule $\rho$ is *applicable* to a branch $\langle \mathcal{S}, \Sigma, \prec \rangle$ if and only if $\mathcal{S}$ contains an instance of the main formulae of $\rho$ and the conditions $\Gamma$ of $\rho$ are satisfied by $\Sigma$ and $\prec$.

$$(\bot)\ \frac{n :: \alpha,\ n :: \neg\alpha}{n :: \bot} \qquad (\neg)\ \frac{n :: \neg\neg\alpha}{n :: \alpha}$$

$$(\land)\ \frac{n :: \alpha \land \beta}{n :: \alpha,\ n :: \beta} \qquad (\lor)\ \frac{n :: \neg(\alpha \land \beta)}{n :: \neg\alpha\ |\ n :: \neg\beta}$$

$$(\Box_i)\ \frac{n :: \Box_i \alpha\ ;\ n \xrightarrow{i} n'}{n' :: \alpha} \qquad (\Diamond_i)\ \frac{n :: \neg\Box_i \alpha}{n'^\star :: \neg\alpha\ ;\ \Gamma'_1\ |\ n'^\star :: \neg\alpha\ ;\ \Gamma'_2}$$

where $\Gamma'_1 = \{n \xrightarrow{i} n'^\star, n'^\star \in \min_{\prec} \Sigma_i(n)\}$ and

$\Gamma'_2 = \{n \xrightarrow{i} n'^\star,\ n \xrightarrow{i} n''^\star,\ n''^\star \prec n'^\star,\ n''^\star \in \min_{\prec} \Sigma_i(n)\}$

$$(\boxtimes_i)\ \frac{n :: \boxtimes_i \alpha\ ;\ n \xrightarrow{i} n',\ n' \in \min_{\prec} \Sigma_i(n)}{n' :: \alpha}$$

$$(\Diamond_i)\ \frac{n :: \neg\boxtimes_i \alpha}{n'^\star :: \neg\alpha\ ;\ n \xrightarrow{i} n'^\star,\ n'^\star \in \min_{\prec} \Sigma_i(n)}$$

**Figure 4: Tableau rules for $\widetilde{\mathcal{L}}$.**

The Boolean rules together with $(\Box_i)$ are as usual and need no explanation. Rule $(\boxtimes_i)$ propagates formulae in the scope of a defeasible necessity operator to the most preferred (with respect to $\prec$) of all accessible nodes. Rule $(\Diamond_i)$ creates a preferred accessible node with the corresponding labeled formulae as content. Rule $(\Diamond_i)$ replaces the standard rule for $\Diamond$-formulae and requires a more thorough explanation. When creating a new accessible node, there are two possibilities: Either (*i*) it is minimal (with respect to $\prec$) amongst all the accessible nodes, in which case the result is the same as that of applying Rule $(\Diamond_i)$, or (*ii*) it is not minimal, in which case there must be a most preferred accessible node that is more preferred (with respect to $\prec$) than the newly created one. (This splitting is of the same nature as that in the ($\vee$)-rule, i.e., it fits the purpose of a proof by cases.)

DEFINITION 12. *A tableau $\mathcal{T}$ for $\alpha \in \widetilde{\mathcal{L}}$ is the limit of a sequence $\mathcal{T}^0, \ldots, \mathcal{T}^n, \ldots$ of sets of branches where the initial $\mathcal{T}^0 = \{\langle \{0 :: \alpha\}, \emptyset, \emptyset \rangle\}$ and every $\mathcal{T}^{i+1}$ is obtained from $\mathcal{T}^i$ by the application of one of the rules in Figure 4 to some branch $\langle \mathcal{S}, \Sigma, \prec \rangle \in \mathcal{T}^i$. Such a limit is denoted $\mathcal{T}^\infty$.*

We make the so-called *fairness assumption*: Any rule that *can* be applied *will* eventually be applied, i.e., the order of rule applications is not relevant. We say a tableau is *saturated* if no rule is applicable to any of its branches.

DEFINITION 13. *A branch $\langle \mathcal{S}, \Sigma, \prec \rangle$ is closed if and only if $n :: \bot \in \mathcal{S}$ for some $n$. A saturated tableau $\mathcal{T}$ for $\alpha \in \widetilde{\mathcal{L}}$ is closed if and only if all its branches are closed. (If $\mathcal{T}$ is not closed, then we say that it is an open tableau.)*

For an example of construction of a tableau, consider the sentence $\alpha = \mathbin{\reflectbox{$\approx$}}(p \to q) \to (\Box p \to \Box q)$ (which is not valid). Figure 5 depicts the (open) tableau for $\neg\alpha = \mathbin{\reflectbox{$\approx$}}\neg(p \wedge \neg q) \wedge \Box p \wedge \neg\Box q$.

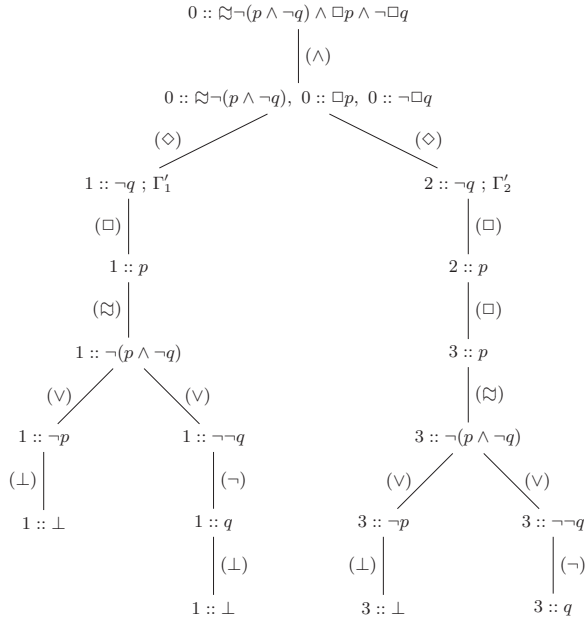

$\Gamma'_1$ = add $(0,1)$ to $\Sigma$ and 1 to $\min_\prec \Sigma(0)$
$\Gamma'_2$ = add $(0,2)$ and $(0,3)$ to $\Sigma$, $(3,2)$ to $\prec$ and 3 to $\min_\prec \Sigma(0)$

**Figure 5: Visualization of an open tableau for the formula $\mathbin{\reflectbox{$\approx$}}\neg(p \wedge \neg q) \wedge \Box p \wedge \neg\Box q$.**

From the open tableau in Figure 5 we extract the preferential Kripke model $\mathscr{P}$ depicted in Figure 6. (In Figure 6 the understanding is that $3 \prec 2$ and that 0 is *incomparable* with respect to $\prec$ to the other possible worlds.)

We are now ready to state the main result of this section.

THEOREM 2. *The tableau calculus for $\widetilde{\mathcal{L}}$ is sound and complete with respect to the modal preferential semantics.*

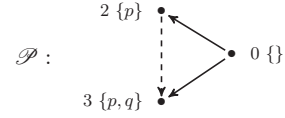

**Figure 6: Preferential Kripke model $\mathscr{P}$ constructed from Figure 5.**

PROOF. See Appendix A.3. □

It can easily be checked that in the construction of the tableau there is only a finite number of distinct states since every formula generated by the application of a rule is a sub-formula of the original one. Hence we have a decision procedure for $\widetilde{\mathcal{L}}$.

We end this section with a brief remark on complexity. It is well-known that satisfiability checking for modal logic K and $\mathsf{K}_n$ are both PSPACE-complete [23, 34]. The addition of $\mathbin{\reflectbox{$\approx$}}$ and $\Diamond\!\!\!\cdot$ to the language does not affect the space complexity of the resulting tableaux. If the formula at the root of the tableau is $\alpha$, and $|\alpha| = m$, then the space requirement for each label is at most $O(m)$. Since there exists a saturated tableau with depth at most $O(m^2)$, the total space requirement is $O(m^3)$.

## 7. ADDING DEFEASIBLE ARGUMENTS

An obvious next step to the work presented here is the integration of $\widetilde{\mathcal{L}}$ with a KLM-style defeasible consequence relation $\mathrel{|\!\sim}$, since this would allow for the expression of both defeasible modalities and defeasible argument forms.[5] First we need some definitions.

Given $\mathscr{P} = \langle W, R, V, \prec \rangle$ and $\alpha, \beta \in \mathcal{L}$, the defeasible statement $\alpha \mathrel{|\!\sim} \beta$ holds in $\mathscr{P}$ (denoted $\mathscr{P} \Vdash \alpha \mathrel{|\!\sim} \beta$) if and only if $\min_\prec [\![\alpha]\!] \subseteq [\![\beta]\!]$, i.e., every $\prec$-minimal $\alpha$-world is a $\beta$-world. As an example, in the model $\mathscr{P}$ of Figure 1, we have $\mathscr{P} \Vdash p \mathrel{|\!\sim} \Box q$ (but note that $\mathscr{P} \not\Vdash p \to \Box q$). We also have $\mathscr{P} \Vdash \neg p \mathrel{|\!\sim} \Diamond(\neg p \wedge \Box q)$ and $\mathscr{P} \not\Vdash \Box\neg q \mathrel{|\!\sim} \neg q$ (from the latter follows $\mathscr{P} \not\Vdash \Box\neg q \to \neg q$).

It is worth noting that if only a classical modal language is assumed, then defeasible statements here still have the same intuition as mentioned in the Introduction. To witness, the statement $\Diamond\alpha \mathrel{|\!\sim} \Box\beta$ just says that "all normal worlds with an $\alpha$-successor have only $\beta$-successors". That is, any $\mathrel{|\!\sim}$-statement still refers only to normality in the premise, or, in this case, of the 'actual' world. In our enriched language we shall be able to make statements of the form $\alpha \mathrel{|\!\sim} \mathbin{\reflectbox{$\approx$}}\beta$.

We say that a preferential Kripke model $\mathscr{P}$ satisfies a set of defeasible statements if each such statement holds in $\mathscr{P}$. Given a set $X$ of defeasible statements, we say that $X$ (preferentially) entails the defeasible statement $\alpha \mathrel{|\!\sim} \beta$ (denoted $X \models \alpha \mathrel{|\!\sim} \beta$) if every preferential model satisfying all the statements in $X$ also satisfies $\alpha \mathrel{|\!\sim} \beta$. (It is easy to see that $\models$ here is exactly the same entailment relation from Definition 6, just restated in terms of $\mathrel{|\!\sim}$-statements.)

We can now relate the truth of $\widetilde{\mathcal{L}}$-sentences in a preferential model with that of defeasible statements, as the following result shows.

---
[5]Here, $\mathrel{|\!\sim}$ need not be a new connective in the language but can rather have the same status as, e.g., *subsumption* and defeasible versions thereof in description logics [1, 7, 9].

LEMMA 2. *Let $\alpha \in \widetilde{\mathcal{L}}$ and $\mathscr{P}$ be a preferential Kripke model. Then $\mathscr{P} \Vdash \alpha$ if and only if $\mathscr{P} \Vdash \neg\alpha \mathrel{\vert\!\sim} \bot$.*

PROOF. See Appendix A.4. □

This result raises the obvious question on whether and how entailment of $\widetilde{\mathcal{L}}$-sentences relates to that of $\mathrel{\vert\!\sim}$-statements.

DEFINITION 14. *Let $\mathcal{K} \subseteq \widetilde{\mathcal{L}}$. $\mathcal{K}^{\mathrel{\vert\!\sim}} := \{\neg\alpha \mathrel{\vert\!\sim} \bot \mid \alpha \in \mathcal{K}\}$.*

THEOREM 3. *$\mathcal{K} \models \alpha$ if and only if $\mathcal{K}^{\mathrel{\vert\!\sim}} \models \neg\alpha \mathrel{\vert\!\sim} \bot$.*

PROOF. See Appendix A.4. □

Hence, preferential entailment in $\widetilde{\mathcal{L}}$ reduces to preferential entailment of $\mathrel{\vert\!\sim}$-statements in the language of $\widetilde{\mathcal{L}}$. Note that soundness of KLM postulates for modal preferential reasoning [8, 10] is preserved when moving from $\mathcal{L}$ to $\widetilde{\mathcal{L}}$. An immediate consequence of this is that the existence of a sound and complete KLM-style $\mathrel{\vert\!\sim}$-based proof system [33] for $\widetilde{\mathcal{L}}$ would define a decision procedure for the extension of $\widetilde{\mathcal{L}}$ with $\mathrel{\vert\!\sim}$. At present we can only conjecture that a proof system along these lines exists, and is based on the integration of the tableau-based proof procedure for $\widetilde{\mathcal{L}}$ presented in Section 6 and the tableau calculi of Giordano et al [20].

## 8. DISCUSSION AND RELATED WORK

To the best of our knowledge, the first attempt to formalize a notion of relative normality in the context of defeasible reasoning was that of Delgrande [17] in which a conditional logic of normality is defined. Given the relationship between the general constructions on which we base our work and those by Kraus et al., most of the remarks in the comparison made by Lehmann and Magidor [35, Section 3.7] are applicable in comparing Delgrande's approach to ours and we do not repeat them here. We note though that, like Kraus et al. and Boutilier, Delgrande focuses on defeasibility of argument forms rather than modes of reasoning as we studied here. Contrary to them, Delgrande adopts the semantics of standard conditional logics [14, Chapter 10], which is based on a (general) selection function picking out the most normal worlds relative to the current one. In his setting, a conditional $\alpha \Rightarrow \beta$ holds at a world $w$ if and only if the set of most normal $\alpha$-worlds (relative to $w$) are also $\beta$-worlds. We can capture Delgrande's conditionals in our approach with $\mathrel{\reflectbox{$\sim$}\mkern-6mu\sim}$-formulae of the form $\mathrel{\reflectbox{$\sim$}\mkern-6mu\sim}(\alpha \to \beta)$ in the class of S5 preferential Kripke models.

Boutilier's expressive conditional logics of normality [5] act as unifying framework for a number of conditional logics, including those of Delgrande and Kraus et al. but do not suffice to define $\mathrel{\reflectbox{$\sim$}\mkern-6mu\sim}$. This is because his modalities are defined directly from a preference order, and do not influence the meaning of any further modalities added to the language.

Baltag and Smets [2] also employ preference orders to refer to the normality of accessible worlds, but their aims and resulting semantics differ from ours in key aspects. They define multi-agent epistemic and doxastic *plausibility models* similar to our preferential Kripke models. Each accessibility relation is induced by a corresponding preference order and linked to an agent whose beliefs are determined by what the agent deems epistemically possible. Minimality, or *doxastic appearance*, is therefore determined relative to an epistemic context, which is induced as an equivalence relation on worlds. This results in modalities of knowledge, (conditional) belief and safe belief that are somewhat related to our defeasible modalities.

In contrast, our work offers a preferential semantic framework independent of a specific application area. We assume (for now) a single preference order across worlds in each Kripke model. The preference order informs the meaning of existing modalities by considering minimality in accessible worlds, where accessibility is determined independently from the preference order. The key difference between our proposal and plausibility models is therefore that our classical modalities are defined independently from any preference order. The special case of a single modality which does correspond to a (connected) preference order yields a logic in which $\mathrel{\reflectbox{$\sim$}\mkern-6mu\sim}$ defines a belief operator. This follows from the conflation of accessibility and preference in plausibility models.

As we have seen, Britz et al. [8, 10] also propose a general semantic framework for preferential modal logics, but they focus on defeasible arguments rather than on defeasible modalities. As such, the semantics introduced there provides a foundation for the semantics of defeasible modalities, but the syntax of preferential modal logic also does not suffice to define preferential modalities such as ours.

Booth et al. [4] introduce an operator with which one can refer directly in the language to those *most typical* situations in which a given sentence is true. For instance, in their enriched language, a sentence of the form $\overline{\alpha}$ refers to the 'most typical' $\alpha$-worlds in a semantics similar to ours. One of the advantages of such an extension is the possibility to make statements of the kind "all normal $\alpha$-worlds are normal $\beta$-worlds", thereby shifting the focus of normality from the antecedent by also allowing us to talk about normality in the consequent. This additional expressivity can also be obtained by the addition of the modality $\square$ of Modular Gödel-Löb logic to express normality syntactically [6, 20]:

$$\overline{\alpha} \equiv_{\text{def}} \square\neg\alpha \wedge \alpha \qquad (9)$$

Despite the gain in expressivity, both these proposals remain propositional in nature in that the only modality allowed is the one with semantics determined by the preference order. Britz et al. extended propositional preferential reasoning to the modal case [8, 10], but the modalities under consideration there remain classical — their meaning remains as in propositional modal logic, despite the underlying preferential semantics of the logic due to the extension of the language with conditional statements of the form $\alpha \mathrel{\vert\!\sim} \beta$.

If we internalize the preference relation as a modality and enrich our classical modal language with converse modalities and nominals [3], then $\mathrel{\reflectbox{$\sim$}\mkern-6mu\sim}$ can be given an entirely classical treatment as follows:

$$\mathrel{\reflectbox{$\sim$}\mkern-6mu\sim}\alpha \equiv_{\text{def}} \bigvee_{o \in \mathcal{O}} (o \wedge \square(\neg\alpha \to \Diamond_\prec(\alpha \wedge \check{\Diamond}o))) \qquad (10)$$

where $\Diamond_\prec$ is the dual of the modality characterizing the preference relation [6], $\check{\Diamond}$ is the converse of $\Diamond$ and $\mathcal{O}$ is a set of nominals. Then $\mathrel{\reflectbox{$\sim$}\mkern-6mu\sim}\alpha$ is true at a world $w$ in a (hybrid) Kripke model if and only if $w$ is the denotation of some nominal $o \in \mathcal{O}$ and every $\neg\alpha$-world that is accessible from $w$ is less normal than some $\alpha$-world which is accessible from $w$. (Of course, besides ensuring that each nominal is interpreted as at most one possible world one also has to make sure that each possible world in a Kripke model is the denotation of some nominal $o \in \mathcal{O}$. This is warranted in the class of *named models* [3, pp. 439–447].)

The definition in (10) above has the inconvenience of requiring infinitary disjunctions [30] in the language. We can replace (10) with an infinitely denumerable collection of axiom schemata given by:

$$(F) \quad @_o \mathbin{\rotatebox[origin=c]{180}{$\approx$}} \alpha \leftrightarrow @_o \Box (\Box_{\prec} \neg \Diamond o \rightarrow \alpha) \qquad (11)$$

As mentioned earlier, making use of such a machinery takes us to a much more expressive language. Note though that complexity-wise we remain in the same class — satisfiability in the basic hybrid logic like the one briefly sketched above is PSPACE-complete [3, Theorem 7.21].

Finally, despite the similarities between the tableau method we presented here and the one by Giordano et al. [20], they remain largely superficial. First, our preferential semantics counts as a proper generalization of the KLM approach to full modal logic, whereas theirs is an embedding of propositional KLM consequence relations in an enriched language. Second, again, in their approach the preference relation is explicit and cast as an additional modality, requiring a special tableau rule to deal with it. Here the preference relation is not present in the language and materializes only in the inner workings of our tableau method.

## 9. CONCLUSION AND FUTURE WORK

The main contribution of the present paper is the provision of a natural, simple and intuitive framework within which to represent defeasible modes of inference. The defeasible modalities we introduced here refer to the relative normality of *accessible worlds*, unlike syntactic characterizations of normality [4, 5, 20, 21], which refer to the relative normality of worlds in which a given sentence is true, or $\mathrel{|\!\sim}$ [33, 35], which refers to the relative normality of the worlds in which the premise is true.

We have seen that the modal logics obtained through the addition of $\mathbin{\rotatebox[origin=c]{180}{$\approx$}}_i$ are monotonic (Theorem 1). Although a logic based on $\widetilde{\mathcal{L}}$ can be extended to include a nonmonotonic conditional $\mathrel{|\!\sim}$, such an extension does not make the addition of $\mathbin{\rotatebox[origin=c]{180}{$\approx$}}_i$ a superfluous extension to the language, since $\mathbin{\rotatebox[origin=c]{180}{$\approx$}}_i$ cannot be expressed in terms of $\mathrel{|\!\sim}$. One avenue for future research is therefore integrating $\mathbin{\rotatebox[origin=c]{180}{$\approx$}}_i$ with our approach to modal preferential reasoning [8, 10], since this would allow for the expression of both defeasible arguments and defeasible modalities. First steps towards this aim were presented in Section 7. Once this is in place, a deeper exploration of applications in various modal logics is warranted.

Here we have investigated the case where a single preference ordering among worlds is assumed. As we have seen, this fits the bill in capturing defeasibility of action effects or obligations, where an 'objective' or commonly agreed upon notion of normality can be quite easily justified. When moving to defeasible notions of knowledge or belief, though, a multi-preference based approach seems to be more appropriate, as agents may have different views on which worlds are more normal than others, i.e., preferences become *subjective* or at least relative to an agent [2].

Here we have investigated defeasible modalities in the system K. Our basic framework paves the way for exploring similar notions of defeasibility and additional properties in specific systems of modal logics. Once this is in place we will be able to investigate further applications of defeasible modalities in e.g. dynamic epistemic logic [36] as well as in other similarly structured logics, such as description logics [1]. We are currently investigating such extensions.

Finally, from a knowledge representation perspective, when one deals with knowledge bases, issues related to modularization [25, 26, 27, 28], knowledge base update and repair [24, 39, 40] as well as knowledge base maintenance and versioning [19] show up. These are tasks acknowledged as important by the community in the classical case [31] and that also make sense in a nonmonotonic setting. When moving to a defeasible approach, though, such tasks have to be reassessed and specific methods and techniques redesigned. This constitutes an avenue worthy of exploration.

## 10. REFERENCES

[1] F. Baader, D. Calvanese, D. McGuinness, D. Nardi, and P. Patel-Schneider, editors. *The Description Logic Handbook: Theory, Implementation and Applications.* Cambridge University Press, 2 edition, 2007.

[2] A. Baltag and S. Smets. A qualitative theory of dynamic interactive belief revision. In G. Bonanno, W. van der Hoek, and M. Wooldridge, editors, *Logic and the Foundations of Game and Decision Theory (LOFT7)*, pages 13–60. Amsterdam Univ. Press, 2008.

[3] P. Blackburn, M. de Rijke, and Y. Venema. *Modal Logic.* Cambridge Tracts in Theoretical Computer Science. Cambridge University Press, 2001.

[4] R. Booth, T. Meyer, and I. Varzinczak. PTL: A propositional typicality logic. In L. Fariñas del Cerro, A. Herzig, and J. Mengin, editors, *Proceedings of the 13th European Conference on Logics in Artificial Intelligence (JELIA)*, number 7519 in LNCS, pages 107–119. Springer, 2012.

[5] C. Boutilier. Conditional logics of normality: A modal approach. *Artificial Intelligence*, 68(1):87–154, 1994.

[6] K. Britz, J. Heidema, and W. Labuschagne. Semantics for dual preferential entailment. *Journal of Philosophical Logic*, 38:433–446, 2009.

[7] K. Britz, J. Heidema, and T. Meyer. Semantic preferential subsumption. In J. Lang and G. Brewka, editors, *Proc. International Conference on Principles of Knowledge Representation and Reasoning (KR)*, pages 476–484. AAAI Press/MIT Press, 2008.

[8] K. Britz, T. Meyer, and I. Varzinczak. Preferential reasoning for modal logics. *Electronic Notes in Theoretical Computer Science*, 278:55–69, 2011.

[9] K. Britz, T. Meyer, and I. Varzinczak. Semantic foundation for preferential description logics. In D. Wang and M. Reynolds, editors, *Proc. Australasian Joint Conference on Artificial Intelligence*, number 7106 in LNAI, pages 491–500. Springer, 2011.

[10] K. Britz, T. Meyer, and I. Varzinczak. Normal modal preferential consequence. In M. Thielscher and D. Zhang, editors, *Proc. Australasian Joint Conference on Artificial Intelligence*, number 7691 in LNAI, pages 505–516. Springer, 2012.

[11] K. Britz and I. Varzinczak. Defeasible modes of inference: A preferential perspective. In *International Workshop on Nonmonotonic Reasoning (NMR)*, 2012.

[12] M. Castilho, O. Gasquet, and A. Herzig. Formalizing action and change in modal logic I: the frame problem. *Journal of Logic and Computation*, 9(5):701–735, 1999.

[13] M. Castilho, A. Herzig, and I. Varzinczak. It depends on the context! A decidable logic of actions and plans based on a ternary dependence relation. In *Intl. Workshop on Nonmonotonic Reasoning (NMR)*, 2002.

[14] B. Chellas. *Modal logic: An introduction*. Cambridge University Press, 1980.

[15] G. Crocco and P. Lamarre. On the connections between nonmonotonic inference systems and conditional logics. In R. Nebel, C. Rich, and W. Swartout, editors, *Proc. International Conference on Principles of Knowledge Representation and Reasoning (KR)*, pages 565–571. Morgan Kaufmann Publishers, 1992.

[16] G. De Giacomo and M. Lenzerini. PDL-based framework for reasoning about actions. In M. Gori and G. Soda, editors, *Proceedings of the 4th Congress of the Italian Association for Artificial Intelligence (IA*AI)*, number 992 in LNAI, pages 103–114. Springer-Verlag, 1995.

[17] J. Delgrande. A first-order logic for prototypical properties. *Artificial Intelligence*, 33:105–130, 1987.

[18] R. Demolombe, A. Herzig, and I. Varzinczak. Regression in modal logic. *Journal of Applied Non-Classical Logics*, 13(2):165–185, 2003.

[19] E. Franconi, T. Meyer, and I. Varzinczak. Semantic diff as the basis for knowledge base versioning. In *International Workshop on Nonmonotonic Reasoning (NMR)*, 2010.

[20] L. Giordano, V. Gliozzi, N. Olivetti, and G. Pozzato. Analytic tableaux calculi for KLM logics of nonmonotonic reasoning. *ACM Transactions on Computational Logic*, 10(3):18:1–18:47, 2009.

[21] L. Giordano, N. Olivetti, V. Gliozzi, and G. Pozzato. $\mathcal{ALC} + T$: a preferential extension of description logics. *Fundamenta Informaticae*, 96(3):341–372, 2009.

[22] R. Goré. Tableau methods for modal and temporal logics. In M. D'Agostino, D. Gabbay, R. Hähnle, and J. Posegga, editors, *Handbook of Tableau Methods*, pages 297–396. Kluwer Academic Publishers, 1999.

[23] J. Halpern and Y. Moses. A guide to completeness and complexity for modal logics of knowledge and belief. *Artificial Intelligence*, 54:319–379, 1992.

[24] A. Herzig, L. Perrussel, and I. Varzinczak. Elaborating domain descriptions. In G. Brewka, S. Coradeschi, A. Perini, and P. Traverso, editors, *Proceedings of the 17th European Conference on Artificial Intelligence (ECAI)*, pages 397–401. IOS Press, 2006.

[25] A. Herzig and I. Varzinczak. Domain descriptions should be modular. In R. López de Mántaras and L. Saitta, editors, *Proceedings of the 16th European Conference on Artificial Intelligence (ECAI)*, pages 348–352. IOS Press, 2004.

[26] A. Herzig and I. Varzinczak. Cohesion, coupling and the meta-theory of actions. In L. Kaelbling and A. Saffiotti, editors, *Proc. International Joint Conference on Artificial Intelligence (IJCAI)*, pages 442–447. Morgan Kaufmann Publishers, 2005.

[27] A. Herzig and I. Varzinczak. On the modularity of theories. In R. Schmidt, I. Pratt-Hartmann, M. Reynolds, and H. Wansing, editors, *Advances in Modal Logic*, 5, pages 93–109. King's College Publications, 2005.

[28] A. Herzig and I. Varzinczak. A modularity approach for a fragment of $\mathcal{ALC}$. In M. Fisher, W. van der Hoek, B. Konev, and A. Lisitsa, editors, *Proceedings of the 10th European Conference on Logics in Artificial Intelligence (JELIA)*, number 4160 in LNAI, pages 216–228. Springer-Verlag, 2006.

[29] A. Herzig and I. Varzinczak. Metatheory of actions: beyond consistency. *Artificial Intelligence*, 171:951–984, 2007.

[30] C. Karp. *Languages with Expressions of Infinite Length*. North-Holland, 1964.

[31] B. Konev, D. Walther, and F. Wolter. The logical difference problem for description logic terminologies. In A. Armando, P. Baumgartner, and G. Dowek, editors, *Proc. International Joint Conference on Automated Reasoning (IJCAR)*, number 5195 in LNAI, pages 259–274. Springer-Verlag, 2008.

[32] M. Kracht and F. Wolter. Properties of independently axiomatizable bimodal logics. *Journal of Symbolic Logic*, 56(4):1469–1485, 1991.

[33] S. Kraus, D. Lehmann, and M. Magidor. Nonmonotonic reasoning, preferential models and cumulative logics. *Artificial Intelligence*, 44:167–207, 1990.

[34] R. Ladner. The computational complexity of provability in systems of modal propositional logic. *SIAM Journal on Computing*, 6(3):467–480, 1977.

[35] D. Lehmann and M. Magidor. What does a conditional knowledge base entail? *Artificial Intelligence*, 55:1–60, 1992.

[36] H. van Ditmarsch, W. van der Hoek, and B. Kooi. *Dynamic Epistemic Logic*. Springer, 2007.

[37] I. Varzinczak. Causalidade e dependência em raciocínio sobre ações ("Causality and dependency in reasoning about actions"). M.Sc. thesis, Universidade Federal do Paraná, Curitiba, Brazil, 2002.

[38] I. Varzinczak. *What is a good domain description? Evaluating and revising action theories in dynamic logic*. PhD thesis, Univ. Paul Sabatier, Toulouse, 2006.

[39] I. Varzinczak. Action theory contraction and minimal change. In J. Lang and G. Brewka, editors, *Proc. International Conference on Principles of Knowledge Representation and Reasoning (KR)*, pages 651–661. AAAI Press/MIT Press, 2008.

[40] I. Varzinczak. On action theory change. *Journal of Artificial Intelligence Research*, 37:189–246, 2010.

[41] D. Zhang and N. Foo. EPDL: A logic for causal reasoning. In B. Nebel, editor, *Proc. International Joint Conference on Artificial Intelligence (IJCAI)*, pages 131–138. Morgan Kaufmann Publishers, 2001.

[42] D. Zhang and N. Foo. Interpolation properties of action logic: Lazy-formalization to the frame problem. In S. Flesca, S. Greco, N. Leone, and G. G. Ianni, editors, *Proceedings of the 8th European Conference on Logics in Artificial Intelligence (JELIA)*, number 2424 in LNCS, pages 357–368. Springer-Verlag, 2002.

[43] D. Zhang and N. Foo. Frame problem in dynamic logic. *Journal of Applied Non-Classical Logics*, 15(2):215–239, 2005.

# APPENDIX

## A. PROOFS OF MAIN RESULTS

### A.1 Proof of Lemma 1

- Proving the 'only if' part: Let $\alpha \in \mathcal{L}$ be such that $\mathscr{M} \Vdash \alpha$, where $\mathscr{M} = \langle W, R, V \rangle$. Then $\mathscr{M}, w \Vdash \alpha$ for every $w \in W$. Let $\mathscr{P} = \langle W, R, V, \prec \rangle$ for some $\prec \subseteq W \times W$. Since $\alpha \in \mathcal{L}$, $\alpha$'s truth conditions do not depend on $\prec$. Then, given that $\alpha$ is true at every $w \in W$, it follows that $[\![\alpha]\!] = W$ and therefore $\mathscr{P} \Vdash \alpha$.

- Proving the 'if' part: Let $\alpha \in \mathcal{L}$ be such that $\mathscr{P} \Vdash \alpha$, where $\mathscr{P} = \langle W, R, V, \prec \rangle$. Then $[\![\alpha]\!] = W$. Since $\alpha \in \mathcal{L}$, it follows that $\mathscr{M}, w \Vdash \alpha$ for every $w \in W$ with $\mathscr{M} = \langle W, R, V \rangle$. Hence $\mathscr{M} \Vdash \alpha$. □

### A.2 Proof of Theorem 1

- Showing Inclusion: Let $\alpha \in \mathcal{K}$. Since every preferential Kripke model of $\mathcal{K}$ is a model of $\alpha$, it immediately follows that $\mathcal{K} \models \alpha$, from which follows $\alpha \in Cn(\mathcal{K})$.

- Showing Idempotency: Let $\alpha \in Cn(\mathcal{K})$. Then $Cn(\mathcal{K}) \models \alpha$ follows by the same argument given for Inclusion above. Hence $\alpha \in Cn(Cn(\mathcal{K}))$. For the other direction, let $\alpha \in Cn(Cn(\mathcal{K}))$. Then $Cn(\mathcal{K}) \models \alpha$. Assume that $\alpha \notin Cn(\mathcal{K})$. Then $\mathcal{K} \not\models \alpha$, and then there exists $\mathscr{P}$ such that $\mathscr{P} \Vdash \mathcal{K}$ but $\mathscr{P} \not\Vdash \alpha$. But from the definition of $Cn(\cdot)$ we have $\mathscr{P} \Vdash Cn(\mathcal{K})$, from which we derive a contradiction. Hence $\alpha \in Cn(\mathcal{K})$.

- Showing Monotonicity: Let $\alpha \in Cn(\mathcal{K}_1)$. Then $\mathcal{K}_1 \models \alpha$. Let $\mathscr{P}$ be such that $\mathscr{P} \Vdash \mathcal{K}_2$. Since $\mathcal{K}_1 \subseteq \mathcal{K}_2$, we have $\mathscr{P} \Vdash \mathcal{K}_1$ too. Hence $\mathscr{P} \Vdash \alpha$ and we have $\mathcal{K}_2 \models \alpha$, and therefore $\alpha \in Cn(\mathcal{K}_2)$. □

### A.3 Proof of Theorem 2

We first show completeness of our tableau method, i.e., if $\alpha \in \widetilde{\mathcal{L}}$ is preferentially valid, then every tableau for $\neg \alpha$ is closed. Equivalently, if there is an open (saturated) tableau for $\alpha$, then $\alpha$ is satisfiable, i.e., there exists a preferential Kripke model $\mathscr{P}$ in which $[\![\alpha]\!] \neq \emptyset$.

In the following, we show that from any open tableau $\mathcal{T}$ for $\alpha \in \widetilde{\mathcal{L}}$ one can construct a preferential Kripke model satisfying $\alpha$, from which the result follows.

Let $\mathcal{T} = \mathcal{T}^\infty$ be an open saturated tableau for the formula $\alpha \in \widetilde{\mathcal{L}}$ (possibly infinite). Then there must be an open branch $\langle \mathcal{S}, \Sigma, \prec \rangle$ in $\mathcal{T}$ (cf. Definition 13). Let the tuple $\mathscr{P}_\mathcal{T} := \langle W_\mathcal{T}, R_\mathcal{T}, V_\mathcal{T}, \prec_\mathcal{T} \rangle$ be defined as follows:

- $W_\mathcal{T} := \{n \mid n :: \beta \in \mathcal{S}\}$;

- $R_\mathcal{T} := \langle R_1, \ldots, R_n \rangle$, where each $R_i := \Sigma(i)$, for $1 \leq i \leq n$;

- $V_\mathcal{T} := v$, where $v : W_\mathcal{T} \times \mathcal{P} \longrightarrow \{0, 1\}$ and $v(n, p) = 1$ if and only if $n :: p \in \mathcal{S}$, and

- $\prec_\mathcal{T} := \prec$.

LEMMA 3. $\mathscr{P}$ *is a preferential Kripke model.*

PROOF. That $\mathscr{M}_\mathcal{T} := \langle W_\mathcal{T}, R_\mathcal{T}, V_\mathcal{T} \rangle$ is a Kripke model follows immediately from the definition of $W_\mathcal{T}$, $R_\mathcal{T}$ and $V_\mathcal{T}$ above. It remains to show that $\prec_\mathcal{T}$ is a strict partial order satisfying the smoothness condition [33]. That is, one has to show that:

- $\prec_\mathcal{T}$ is irreflexive and transitive: This follows from the construction of $\prec$ in Rules $(\diamondsuit_i)$ and $(\Diamond_i)$, since $(i)$ no pair $(n, n)$ is ever added to $\prec$ and $(ii)$ no chain of length greater than 2 is ever added to the preference structure.

- $\prec_\mathcal{T}$ has no infinitely descending chains: Clearly no pair $(n, n')$ is added to $\prec$ beyond those added by Rules $(\diamondsuit_i)$ and $(\Diamond_i)$. Given this one can easily check that $\prec$ must have a minimum.

□

It remains to show that $\mathscr{P}$ above satisfies $\alpha$.

LEMMA 4. *Let $\mathscr{P} = \langle W_\mathcal{T}, R_\mathcal{T}, V_\mathcal{T}, \prec_\mathcal{T} \rangle$ and let $\beta$ be a sub-formula of $\alpha$. If $n :: \beta \in \mathcal{S}$, then $n \in [\![\beta]\!]$.*

PROOF. The proof is by structural induction on the number of connectives in $\beta$.
Base case: $\beta$ is a literal. We have two cases: $(i)$ $\beta = p \in \mathcal{P}$. Then $n :: p \in \mathcal{S}$ if and only if $v(n, p) = 1$ if and only if $V_\mathcal{T}(n, p) = 1$ if and only if $n \in [\![p]\!] = [\![\beta]\!]$. $(ii)$ $\beta = \neg p$ for some $p \in \mathcal{P}$. Then $n :: \neg p \in \mathcal{S}$, and therefore $n :: p \notin \mathcal{S}$, otherwise $n :: \bot \in \mathcal{S}$ (as $\mathcal{T}$ is saturated), contradicting the assumption that $\langle \mathcal{S}, \Sigma, \prec \rangle$ is open. Hence $v(n, p) = 0$, and then $n \notin [\![p]\!]$, from which follows $n \in W_\mathcal{T} \setminus [\![p]\!] = [\![\neg p]\!] = [\![\beta]\!]$.
Induction step: The Boolean cases are as usual. We analyze the modal cases (below $\mathscr{M}_\mathcal{T} = \langle W_\mathcal{T}, R_\mathcal{T}, V_\mathcal{T} \rangle$):

- $\beta = \Box_i \gamma$: If $n :: \Box_i \gamma \in \mathcal{S}$, then $n' :: \gamma \in \mathcal{S}$ by Rule $(\Box_i)$, for every $n'$ such that $(n, n') \in R_i$. By the induction hypothesis, $n' \in [\![\gamma]\!]$ for every $n'$ such that $(n, n') \in R_i$, i.e., $\mathscr{M}_\mathcal{T}, n' \Vdash \gamma$ for every $n'$ such that $(n, n') \in R_i$. From this we conclude $\mathscr{M}_\mathcal{T}, n \Vdash \Box_i \gamma$ and therefore $n \in [\![\Box_i \gamma]\!]$.

- $\beta = \neg \Box_i \gamma$: If $n :: \neg \Box_i \gamma \in \mathcal{S}$, then by Rule $(\diamondsuit_i)$ there exists $n'$ such that $(n, n') \in R_i$ and $n' :: \neg \gamma \in \mathcal{S}$. Then there exists $n'$ such that $(n, n') \in R_i$ and $n' \in [\![\neg \gamma]\!]$, by the induction hypothesis. Hence $n \in [\![\neg \Box_i \gamma]\!]$.

- $\beta = \mathrel{\vcenter{\hbox{$\approx$}}}_i \gamma$: If $n :: \mathrel{\vcenter{\hbox{$\approx$}}}_i \gamma \in \mathcal{S}$, then $n' :: \gamma \in \mathcal{S}$ by Rule $(\mathrel{\vcenter{\hbox{$\approx$}}}_i)$, for every $n'$ such that $n' \in \min_{\prec_\mathcal{T}} R_i(n)$. By the induction hypothesis, $n' \in [\![\gamma]\!]$ for every $n'$ such that $n' \in \min_{\prec_\mathcal{T}} R_i(n)$, and therefore $n \in [\![\mathrel{\vcenter{\hbox{$\approx$}}}_i \gamma]\!]$.

- $\beta = \neg \mathrel{\vcenter{\hbox{$\approx$}}}_i \gamma$: If $n :: \neg \mathrel{\vcenter{\hbox{$\approx$}}}_i \gamma \in \mathcal{S}$, then by Rule $(\Diamond_i)$ there exists $n'$ such that $n' \in \min_{\prec_\mathcal{T}} R_i(n)$ and $n' :: \neg \gamma \in \mathcal{S}$. Then there exists $n'$ such that $n' \in \min_{\prec_\mathcal{T}} R_i(n)$ and $n' \in [\![\neg \gamma]\!]$, by the induction hypothesis. Hence $n \in [\![\neg \mathrel{\vcenter{\hbox{$\approx$}}}_i \gamma]\!]$.

□

Now, since $0 :: \alpha \in \mathcal{S}$, from Lemma 4 we conclude that $0 \in [\![\alpha]\!]$. Hence $[\![\alpha]\!] \neq \emptyset$ for the preferential Kripke model constructed as above, and therefore $\alpha$ is satisfiable, as we wanted to show. □

In the following we show soundness, i.e., if $\alpha \in \widetilde{\mathcal{L}}$ is (preferentially) satisfiable, then there is an open tableau for $\alpha$. Equivalently, if all the tableaux for $\alpha$ are closed, then $\alpha$ is unsatisfiable, i.e., $\neg \alpha$ is valid.

DEFINITION 15. *Let $\mathcal{S}$ be a set of labeled formulae. $\mathcal{S}(n) := \{\beta \mid n :: \beta \in \mathcal{S}\}$.*

DEFINITION 16. $\widehat{\mathcal{S}(n)} := \bigwedge \{\beta \mid \beta \in \mathcal{S}(n)\}$.

LEMMA 5. *If, for every tableau rule that can be applied to $\mathcal{T}^j = \{\ldots, \langle \mathcal{S}^j, \Sigma^j, \prec^j \rangle, \ldots\}$ to produce $\mathcal{T}^{j+1} = \{\ldots, \langle \mathcal{S}^{j+1}, \Sigma^{j+1}, \prec^{j+1} \rangle, \ldots\}$ and for every branch $\langle \mathcal{S}^j, \Sigma^j, \prec^j \rangle \in \mathcal{T}^j$ there exists $n$ such that $\widehat{\mathcal{S}^{j+1}(n)}$ is unsatisfiable, then $\widehat{\mathcal{S}^j(n)}$ is unsatisfiable.*

PROOF. We suffice with the cases of Rules $(\varnothing_i)$ and $(\diamond_i)$.

- Rule $(\varnothing_i)$: If $\mathcal{S}^j$ contains $n :: \neg \varnothing_i \beta$, then an application of Rule $(\varnothing_i)$ creates a new label $n'$, adds $n \xrightarrow{i} n'$ to $\Sigma^j(i)$ to obtain $\Sigma^{j+1}(i)$, adds $n' :: \neg \beta$ to $\mathcal{S}^j$ to obtain $\mathcal{S}^{j+1}$, and sets $n'$ as a minimum in $\Sigma^{j+1}(i)$ with respect to $\prec^{j+1}$ (which extends $\prec^j$). Now, suppose $\widehat{\mathcal{S}^j(n)}$ is satisfiable, but $\widehat{\mathcal{S}^{j+1}(n')}$ is unsatisfiable. Since $\widehat{\mathcal{S}^{j+1}(n')} = \neg \beta$ (as $\mathcal{S}^{j+1}$ is the singleton $\{n' :: \neg \beta\}$ — $n'$ the freshly added label), then $\neg \beta$ must be unsatisfiable, i.e., $\models \beta$. From this and normal necessitation — Rule (6) —, we have $\models \varnothing_i \beta$. Hence $\widehat{\mathcal{S}^j(n)}$ is unsatisfiable too because $n :: \neg \varnothing_i \beta \in \mathcal{S}^j$.

- Rule $(\diamond_i)$: If $\mathcal{S}^j$ contains $n :: \neg \Box_i \beta$, then an application of Rule $(\diamond_i)$ will create a new label $n'$ and either (i) add $n \xrightarrow{i} n'$ to $\Sigma^j(i)$ to obtain $\Sigma^{j+1}(i)$, add $n' :: \neg \beta$ to $\mathcal{S}^j$ to obtain $\mathcal{S}^{j+1}$, and set $n'$ as a minimum in $\Sigma^{j+1}(i)$ with respect to $\prec^{j+1}$ (thereby extending $\prec^j$) or (ii) add $n \xrightarrow{i} n'$ to $\Sigma^j(i)$ to obtain $\Sigma^{j+1}(i)$, add $n' :: \neg \beta$ to $\mathcal{S}^j$ to obtain $\mathcal{S}^{j+1}$, create a new label $n''$ and also add $n \xrightarrow{i} n''$ to $\Sigma^{j+1}(i)$, add $(n'', n')$ to $\prec^j$ to obtain $\prec^{j+1}$ and set $n''$ as a minimum in $\Sigma^{j+1}(i)$ with respect to $\prec^{j+1}$. If (i) is the case, then we have the same argument as for Rule $(\varnothing_i)$ above. Let us assume (ii) is the case. Suppose $\widehat{\mathcal{S}^j(n)}$ is satisfiable, but either $\widehat{\mathcal{S}^{j+1}(n')}$ is unsatisfiable or $\widehat{\mathcal{S}^{j+1}(n'')}$ is unsatisfiable. If $\widehat{\mathcal{S}^{j+1}(n')}$ is unsatisfiable, since $\widehat{\mathcal{S}^{j+1}(n')} = \neg \beta$ we have the same argument as for Rule $(\varnothing_i)$ above. If $\widehat{\mathcal{S}^{j+1}(n'')}$ is unsatisfiable, then since $\widehat{\mathcal{S}^{j+1}(n'')} = \top$, we have $\models \bot$, which implies $\models \Box_i \bot$, and then $\models \Box_i \beta$. Hence $\widehat{\mathcal{S}^j(n)}$ is unsatisfiable too because $n :: \neg \Box_i \beta \in \mathcal{S}^j$.

$\square$

From Lemma 5 we conclude that if all tableaux for $\alpha$ are closed, then every $\widehat{\mathcal{S}(n)}$ is unsatisfiable. In particular $\widehat{\mathcal{S}(0)} = \alpha$ is unsatisfiable. Hence all rules preserve satisfiability when transforming one set of branches into another. This warrants soundness of our tableau rules. $\square$

### A.4 Proofs of Lemma 2 and Theorem 3

**Lemma 2**: Let $\mathscr{P} = \langle W, R, V, \prec \rangle$. $\mathscr{P} \Vdash \alpha$ if and only if $[\![\alpha]\!] = W$ if and only if $[\![\neg \alpha]\!] = \emptyset$ if and only if $\min_\prec [\![\neg \alpha]\!] = \emptyset$ if and only if $\min_\prec [\![\neg \alpha]\!] \subseteq [\![\bot]\!]$ if and only if $\mathscr{P} \neg \alpha \mathrel{\vert\!\sim} \bot$. $\square$

**Theorem 3**: Let $\mathcal{K}^{\mathrel{\vert\!\sim}}$ be obtained from $\mathcal{K}$ as in Definition 14. For the 'only if' part, let $\mathscr{P}$ be such that $\mathscr{P} \Vdash \mathcal{K}^{\mathrel{\vert\!\sim}}$, i.e., $\mathscr{P} \Vdash \neg \beta \mathrel{\vert\!\sim} \bot$ for every $\neg \beta \mathrel{\vert\!\sim} \bot$ in $\mathcal{K}^{\mathrel{\vert\!\sim}}$. From Lemma 2, this is the case if and only if $\mathscr{P} \Vdash \beta$ for every $\beta \in \mathcal{K}$. Hence $\mathscr{P} \Vdash \mathcal{K}$, and since $\mathcal{K} \models \alpha$, we have $\mathscr{P} \Vdash \alpha$ too. From Lemma 2 again we get $\mathscr{P} \Vdash \neg \alpha \mathrel{\vert\!\sim} \bot$. Now, for the 'if' part, let $\mathscr{P}$ be such that $\mathscr{P} \Vdash \mathcal{K}$, i.e., $\mathscr{P} \Vdash \beta$ for all $\beta \in \mathcal{K}$. From Lemma 2, it follows that $\mathscr{P} \Vdash \neg \beta \mathrel{\vert\!\sim} \bot$ for every $\beta \in \mathcal{K}$, and then $\mathscr{P} \Vdash \mathcal{K}^{\mathrel{\vert\!\sim}}$. From this and $\mathcal{K}^{\mathrel{\vert\!\sim}} \models \neg \alpha \mathrel{\vert\!\sim} \bot$ we have $\mathscr{P} \Vdash \neg \alpha \mathrel{\vert\!\sim} \bot$, and therefore by Lemma 2 again we get $\mathscr{P} \Vdash \alpha$. $\square$

## Acknowledgments

The authors are grateful to the anonymous referees for their constructive and useful remarks.

This work is based upon research supported by the National Research Foundation (NRF). Any opinion, findings and conclusions or recommendations expressed in this material are those of the authors and therefore the NRF do not accept any liability in regard thereto. This work was partially funded by Project # 247601, Net2: Network for Enabling Networked Knowledge, from the FP7-PEOPLE-2009-IRSES call.